\documentclass[%
 reprint,
 amsmath,amssymb,
 aps,
]{revtex4-1}

\usepackage{graphicx}
\usepackage{dcolumn}
\usepackage{bm}

\usepackage{comment}

\usepackage{color}

\newtheorem{theorem}{Theorem}

\newtheorem{proposition}[theorem]{Proposition}

\def\Tr{\textnormal{Tr}}

\begin{document}

\preprint{APS/123-QED}

\title{An Elementary Proof of Private Random Number  Generation  \\
from Bell Inequalities}

\author{Carl A.~Miller}
\affiliation{National Institute of Standards and Technology, 100 Bureau Dr., Gaithersburg, MD  20899, USA \\ Joint Center for Quantum Information and Computer Science, 3100 Atlantic Bldg, University of Maryland, College Park, MD  20742, USA} 

\date{\today}

\begin{abstract}
The field of device-independent quantum cryptography has seen enormous success in the past several years,
including security proofs for key distribution and random number generation
that account for arbitrary imperfections in
the devices used.  Full security proofs in the field so far are long and technically deep.  In this paper we show
that the concept of the \textit{mirror adversary} can be used to simplify device-independent proofs.  
We give a short proof that any bipartite Bell violation can be used to generate private random numbers. The proof 
is based on elementary techniques and is self-contained.
\end{abstract}

\pacs{Valid PACS appear here}
\maketitle


Quantum cryptography is based on, among other physical principles, the concept
of \textit{intrinsic randomness}: certain quantum measurements are unpredictable,
even to an adversary who has complete information about the protocol and the apparatus used.
This intrinsic randomness allows a user to generate cryptographic keys that are guaranteed to be secure without the need
for computational assumptions.

Device-independent quantum cryptography is based on a more specific observation:
two or more devices that exhibit superclassical probability correlations (when blocked
from communicating) must be making quantum measurements,
and therefore must be exhibiting random behavior.  This
allows the generation of random numbers even when the devices themselves are not trusted.  This idea has been used in multiple cryptographic
contexts, including randomness expansion and amplification \cite{colbeck2007quantum, Colbeck:2012},
key distribution \cite{Ekert:1991},  and coin-flipping \cite{Silman:2011}, and
has been realized in experiments \cite{pironio2010random, bierhorst2018experimentally}.

Despite the simplicity of the central idea, proofs for device-independent quantum cryptography are challenging
and took several years to develop.  One of the central challenges is proving that 
the random numbers generated by a Bell experiment are secure
even in the presence of quantum side information.  (This level of security is necessary
for quantum key distribution, and also for random number generation if one wishes to use the random numbers
as inputs to another quantum protocol.)
While classical statistical arguments can be used to show
that the outputs of a Bell violation are unpredictable to a classical adversary (see, e.g., \cite{fehr2013security, pironio2013security}) these proofs do not carry over to the case of quantum side information because of the notion of information locking
\cite{DiV:2004}.

Known proofs of Bell randomness in the presence of quantum side information have used tools
that are specific to the context: \cite{vazirani2012, vazirani2014fully} uses reconstruction
properties of quantum-proof randomness extractors, and \cite{MY14-1, miller2017universal,
Dupuis:2016, Arnon:2016, arnon2018practical} are based on inductive arguments centered
on the quantum Renyi divergence function.  Such proofs are long and mathematically complex.  The  recent
paper \cite{Arnon:2016} provides an easily adaptable framework for proving
new results on randomness generation, but it is based on the entropy accumulation theorem \cite{Dupuis:2016},
the proof of which is technically deep.

The goal of the current paper is to provide a compact security proof of Bell randomness in the presence
of quantum side information.
The proof is based on the  \textit{mirror adversary} approach, which uses the fact that an adversary who simply ``mirrors'' the behavior of the
devices is almost as good as an optimal adversary.   The mirror adversary can be considered
as a participant in a larger repeated Bell experiment.
This idea was discussed in a previous paper by the author \cite{Jain:2017}, 
and is a reframing of the commonly used idea of pretty good measurements (see expression
(\ref{pgp}) below).

In the current paper the mirror adversary technique is combined with techniques drawn from other sources
\cite{Tomamichel:2011, Kempe:2011} to give a compact proof of private random number generation from Bell experiments.
(The paper does not attempt to maximize the performance parameters, which are suboptimal compared
to \cite{vazirani2012, vazirani2014fully, MY14-1, miller2017universal,
Dupuis:2016, Arnon:2016, arnon2018practical}.)
The proof is self-contained, with
material from other sources reproved as needed.  The only assertions taken for granted in the proof are Azuma's inequality
(see Theorem 7.2.1 in \cite{Alon:2004}) and Holder's inequality (see Corollary IV.2.6 in \cite{Bhatia:1997}).

The main result is the following (see Theorem~\ref{mainthm}).

\begin{theorem}[Informal]
Suppose that two untrusted devices exhibit a Bell violation of $\delta > 0$ over $N$ rounds.  Then,
$\Omega ( N \delta^6)$ private random bits can be extracted from the outputs of the devices in polynomial
time, using $O ( N )$ bits of public randomness (that is, randomness known to the adversary but not the devices).  The resulting private bits are secure against quantum side information.
\end{theorem}

The mirror adversary technique is a general way of reducing security questions in the quantum context
to classical statistical statements, and it is potentially useful for any cryptographic task
in which security must be proved against a passive entangled adversary.

The author thanks Honghao Fu, Yi-Kai Liu, Ray Perlner, and Thomas Vidick for comments on this paper.  This work is
an official contribution of the U. S. National Institute of Standards and Technology, and is not subject to copyright in the United States.

\paragraph*{\textbf{Preliminaries.}} Throughout the paper, a \textit{register} $Z$ is a finite-dimensional
Hilbert space with a fixed orthonormal basis (the elements of which we call \textit{basic states}).  
A \textit{state} $\phi$ of $Z$ is a density
operator on $Z$.  
Let $\left| Z \right| = \textnormal{dim} ( Z )$.
If $Z = Z_1 \otimes Z_2$ (which we may abbreviate as $Z = Z_1 Z_2$) we will write $\phi^{Z_1}$
for $\Tr_{Z_2} \phi$.  If $Z_2$ is a register and $e$ is a basic
state of $Z_2$, then $\phi^{Z_1}_e$ denotes $\Tr_{Z_2} [ \phi ( I_{Z_1} \otimes
\left| e \right> \left< e \right|) ]$.
As a convenience, if $X$ is an operator on $Z$
and $Y$ is an operator on $Z_1$, then the expression $XY$
means $X ( Y \otimes I_{Z_2 } )$ and the expression $YX$ means
$( Y \otimes I_{Z_2 } )X$.

We give a formalism for nonlocal games and the quantum strategies used in such games.
We begin by formalizing measurements.
An ($N$-fold) measurement strategy on a register $Q$ is a family of 
positive operator-valued measures on $Q$ of the form
\begin{eqnarray}
\left\{ \left\{ F_{\mathbf{u}}^{\mathbf{t}}
\right\}_{\mathbf{t} \in \mathcal{T}^N}
\right\}_{\mathbf{u} \in \mathcal{U}^N},
\end{eqnarray}
where $\mathcal{T}$ and $\mathcal{U}$ are finite sets.  Such a strategy
is \textit{sequential} if for any $t_1, \ldots, t_i \in \mathcal{T}$
and $\mathbf{u} \in \mathcal{U}^N$, the operator
\begin{eqnarray}
F_{\mathbf{u}}^{t_1 \cdots t_i} & := & \sum_{t_{i+1} \cdots t_n}
F_{\mathbf{u}}^{t_1 \cdots t_i t_{i+1} \cdots t_n}
\end{eqnarray}
is independent of the values of $u_{t+1} \cdots u_n$.  (In such a case we can 
simply write $F_{u_1 \cdots u_i}^{t_1 \cdots t_i}$ for $F_{\mathbf{u}}^{t_1 \cdots t_i}$.)
Sequential measurements model the behavior of a quantum player who
receives inputs $u_1, \ldots, u_N$ and produces outputs $t_1, \ldots, t_N$ in sequence.
In such a case, for any $u_1, \ldots, u_i$ and $t_1, \ldots, t_i$ for which
$F_{u_1 \cdots u_i}^{t_1 \cdots t_i} \neq 0$, there is a
$1$-fold measurement strategy on $Q$ given by
\begin{eqnarray*}
\left\{ \left\{ \left( F_{u_1 \cdots u_i}^{t_1 \cdots t_i} \right)^{-1/2}
F_{u_1 \cdots u_{i+1}}^{t_1 \cdots t_{i+1}}
\left( F_{u_1 \cdots u_i}^{t_1 \cdots t_i} \right)^{-1/2}  \right\}_{t_{i+1}} \right\}_{u_{i+1}},
\end{eqnarray*}
which defines the behavior of the player on the $(i+1)$st round conditioned 
on the inputs sequence $u_1, \ldots, u_i$ and output sequence $t_1, \ldots, t_i$
for the first $i$ rounds.  We call these the \textit{conditional} measurement strategies induced
by $\{ \{ F_{\mathbf{u}}^\mathbf{t} \}_{\mathbf{t}} \}_{\mathbf{u}}$.

An \textit{$r$-player nonlocal game} $H$ consists of the following data:
(1) a finite set of input strings $\mathcal{I} = \mathcal{I}_1 \times \cdots \times \mathcal{I}_r$
and a finite set of outputs strings $\mathcal{O} = \mathcal{O}_1 \times \cdots \times \mathcal{O}_r$ 
(2) a probability distribution $p$
on $\mathcal{I}$, and (3) a scoring
function $L \colon \mathcal{I} \times \mathcal{O} \to
\mathbb{R}$.  For such a game, $H^N$ denotes the $N$-fold direct product of $H$
(i.e., the game played $N$ times in parallel, with independently chosen inputs,
and where the score is the sum of scores achieved on each of the $N$ copies of the game).

A measurement strategy for $H$ on a register $Q$ is a measurement strategy on $Q$ of the
form $\{ \{ F_i^o \}_{o \in \mathcal{O}} \}_{i \in \mathcal{I}}$.  Such a strategy is $n$-partite if 
$Q = Q_1 \otimes \cdots \otimes Q_n$ and
\begin{eqnarray}
F_i^o & = & F_{1,i_1}^{o_1} \otimes \cdots \otimes F_{n, i_n}^{o_n}
\end{eqnarray}
where $\{ \{ F_{k, i_k}^{o_k} \}_{o_k \in \mathcal{O}_k} \}_{i_k \in \mathcal{I}_k}$ are measurement
strategies on $Q_k$ for $k = 1, 2, \ldots, n$.  A sequential measurement strategy for the game $H^N$
is an \textit{$n$-partite sequential measurement strategy} if all of its conditional strategies are $n$-partite.  
(This class of strategies models the behavior of players who must play the different rounds of the game in sequence,
and who can communicate in between but not during rounds.)

If $\mathbf{F}$ is a strategy on a register $Q$, and $\phi$ is a state of $Q$, 
then we refer to the pair $(\mathbf{F}, \phi)$ simply as a (quantum) strategy for $Q$.
Let $\omega ( H )$ denote the supremum of the expected score
at $G$ among all quantum strategies.

\begin{proposition}
\label{prop:rep}
Let $H$ be an $r$-player nonlocal game whose scoring
function has range $[-K , K ]$, and let $(\mathbf{F}, \phi)$ be an $n$-partite
sequential measurement strategy for $H^N$.  Then, the probability that the score
achieved by $(\mathbf{F}, \phi)$ exceeds $(\omega ( H ) + \delta )N$ is no more than
\begin{eqnarray}
\label{eupperbound}
e^{-N \delta^2/8 K^2}.
\end{eqnarray}
\end{proposition}

\textit{Proof.}
For each $i = 1 , 2, \ldots, N$, let $W_i$ denote the score
achieved on the $i$th round, and let
\begin{eqnarray}
\overline{W}_i & = & E [ W_i \mid W_{i-1} \cdots W_1 ].
\end{eqnarray}
The sequence $\left( \sum_{j=1}^i ( W_i - \overline{W}_i ) \right)_{i=1}^N$
forms a Martingale, and thus by Azuma's inequality (noting
that $\left| W_i - \overline{W}_i \right| \leq 2K$) the probability that
\begin{eqnarray}
\sum_{i=1}^N ( W_i - \overline{W}_i ) & > & \delta N 
\end{eqnarray}
is upper bounded by (\ref{eupperbound}).  Since $\overline{W}_i \leq w ( H )$, the desired
result follows. $\Box$

For convenience, we also make the following definition.
A \textit{Bell game} is a game $G$ for which we make the following assumptions:
\begin{enumerate}
\item The
input alphabets and output alphabets are all equal to $\{0, 1, 2, \ldots, n-1 \}$ for some $n$.  (We call
$n$ the ``alphabet size.'')
\item The input distribution is uniform.
\item The range of the scoring function is $[-1, 1]$.
\item The optimal classical score is $0$.
\end{enumerate}
Note that any Bell inequality can be put into this form (by an appropriate
affine transformation of the scoring function).

\begin{figure}
\fbox{\parbox{3.2in}{
\textit{Parameters:}
\begin{enumerate}
\item[-] A $2$-player Bell game $G$ with alphabet size $n$.
\item[-] A real number $\delta > 0$ (the degree of Bell violation).
\item[-] Positive integers $N$ (the number of rounds), and  $J$ (the output size).
\end{enumerate}

\begin{enumerate}
\item A pure tripartite state $ABE$ is prepared by Eve, and with $A$
possessed by Alice, $B$ possessed by Bob, and $E$
possessed by Eve.

\item The referee generates uniformly random numbers
$x_1, y_1 \in \{ 1, 2, \ldots, n \}$, gives them as input to Alice and
Bob, respectively, who return outputs $s_1, t_1$.  This is repeated
$(N-1)$ times to obtain input sequences $x_1, \ldots, x_N$, $y_1, \ldots, y_N$
and output sequences $s_1, \ldots, s_N$, $t_1, \ldots, t_N$.  

\item The referee checks whether the average score exceeds
$\delta$.  If not, the protocol is aborted.

\item \label{step:randext}  Let $D$ be a $2$-universal
hash family from $\mathcal{S}^N$ to $\mathbb{F}_2^J$ with $\left| D \right| \leq 4 \left| \mathcal{S}^N \right|^2$
(see subsection~\ref{subsec:hash} in the appendix).  The referee chooses
$F \in D$ at random and outputs $F ( \mathbf{s} )$.
\end{enumerate}
}
}
\caption{The random number generation protocol.}
\label{fig:prot}
\end{figure}

\paragraph*{\textbf{The mirror adversary.}}

\label{sec:mirror}

If $\alpha$ is a quantum-classical state of a
register $QC$ (that is, a state of the form $\sum_c \alpha_c \otimes \left| c \right> \left< c \right|$)
 then the \textit{pretty good measurement}
induced by $\alpha$ on $Q$ is the $C$-valued measurement given by
\begin{eqnarray}
\label{pgp}
\{ (\alpha^Q)^{-1/2} \alpha_c^Q (\alpha^Q)^{-1/2} \}_{c \in C}.
\end{eqnarray}
This is a common construction.  In the cryptographic context
it can be thought of as a ``pretty good'' attempt by an adversary to
use to $Q$ to guess $C$.

Let $\rho$ be a state of the register $Q$.  Then, we can
construct a purification for $\rho$ as follows: let $Q'$ be
an isomorphic copy of $Q$, let $\Phi = \sum_e e \otimes e \in Q \otimes Q'$, where the sum
is over all basic states of $Q$.  Let $\hat{\rho}$ denote
the projector onto the one-dimensional subspace of $Q \otimes Q'$
spanned by $( \sqrt{\rho} \otimes I_{Q'} ) \Phi$.  We call $\hat{\rho}$
the \textit{canonical purification} of $\rho$.  Note that
$\Tr_{Q'} \hat{\rho} = \rho$ while $\Tr_{Q} \hat{\rho} = \rho^\top = \overline{\rho}$.  

The following proposition implies that
a ``pretty good'' adversary in a Bell experiment
simply mirrors the device's measurements.  (As a consequence,
if the devices' measurement were sequential, so are the adversary's.)

\begin{proposition}
\label{prop:whatmirroris}
Let $\rho$ be a state of a register $Q$, and let $\hat{\rho}$ be the canonical
purification of $\rho$ (a state of the registers $QQ'$).
Let $\alpha$ be the state $QC$ that arises
from $\hat{\rho}$ by performing a measurement $\{ R_c \}_{c \in \mathcal{C}}$ on $Q'$ and storing
the result in a register $C$.  Then,
the pretty good measurement induced by $\alpha$ on $Q$
is isomorphic to $\{ \overline{R}_c \}_{c \in \mathcal{C}}$.
\end{proposition}

\textit{Proof.}
The state $\alpha$ is given by the expression $\alpha =  \sum_c \left| c \right> \left< c \right| \otimes \sqrt{\rho} \overline{R_c} \sqrt{\rho}$,
and $\alpha^Q = \rho$. 
The pretty good measurement induced by $\alpha$
on $Q$ is thus isomorphic to $\{ \rho^{-1/2} \sqrt{\rho} \overline{R_c}
\sqrt{\rho} \rho^{-1/2} \} = \{ \overline{R_c} \}$.  $\Box$

The next proposition, which is a modification
of a result from \cite{Tomamichel:2011}, 
expresses the fact that if the pretty good measurement yields (almost) no
information about a classical register $C$, then that register must be (almost) uniformly random.
We state a version 
that will be useful in the device-independent context.
Let $Z$ denote a classical
register with two basic states, $abort$ and $succ$.  

\begin{proposition}
\label{prop:pgp}
Let $\alpha$ be a state of a tripartite register $QCZ$ which is classical on $CZ$.
Let $\{ R_{z } \}$ and $\{ R_{cz} \}$ denote the pretty good measurements on $Q$:
\begin{eqnarray}
R_{cz} & = & (\alpha^{Q} )^{-1/2} \alpha^Q_{cz} (\alpha^{Q} )^{-1/2}, \\
R_{z} & = & (\alpha^{Q} )^{-1/2} \alpha^Q_{z} (\alpha^{Q} )^{-1/2}.
\end{eqnarray}
Let $f  =  \Tr [ \alpha^Q_{succ} R_{succ} ]$
and
\begin{eqnarray}
f' & = & \sum_c \Tr [ \alpha^Q_{c, succ} R_{c, succ } ].
\end{eqnarray}
Then,
\begin{eqnarray}
\left\| \alpha_{succ}^{QC} - \alpha_{succ}^{Q} \otimes U_C \right\|_1 
& \leq & \sqrt{ f' \left| C \right| - f},
\label{succapprox}
\end{eqnarray}
where $U_C$ denotes the completely mixed state on $C$.
\end{proposition}

The proof is given in the appendix.
Note that the quantity $f$ is the probability of the event that both $Z = succ$ and
that an adversary who uses the pretty good measurement will guess
that $Z = succ$.  The quantity $f'$ is the probability that the previous event
holds \textit{and} the adversary guesses $C$.  If $f' = f/ \left| C \right|$ (that is,
if the adversary's guess at $C$ is no better than random)
then the term on the right side of (\ref{succapprox}) is equal to zero.

\paragraph*{\textbf{Guessing games.}} The following is roughly the same as the construction of immunization games in \cite{Kempe:2011}.
Let $G = ( (\mathcal{X}, \mathcal{Y}), (\mathcal{S}, \mathcal{T}), p, L)$ be a $2$-player
Bell game with alphabet size $n$, and let $K > 0$.
Then we define a new $3$-player game $G_K$ as follows.
\begin{enumerate}
\item The input alphabets for the three players are $\mathcal{X}, \mathcal{Y}$ and $\mathcal{X} \times
\mathcal{Y}$, respectively,
and the output alphabets are $\mathcal{S}, \mathcal{T}$ and $\mathcal{S}$, respectively.
\item The probability distribution is uniform on triples of the form $(x, y, (x, y))$, with $x \in
\mathcal{X}, y \in \mathcal{Y}$.
\item The score assigned to an input triple $(x,y, (x,y))$ and output triple $(s, t, s')$
is $L ( x, y, s, t)$ if $s = s'$, and is $(-K)$ otherwise.
\end{enumerate}

\begin{proposition}
\label{prop:limiting}
For any Bell game $G$ with alphabet size $n$, $\omega( G_K ) \leq  4 n/\sqrt{ K}$.
\end{proposition}

Our proof is similar to \cite{Kempe:2011}.  We will use the process
described in Figure~\ref{fig:fclass}.

\begin{figure}
\fbox{\parbox{3.2in}{
Let $G$ be a Bell game and $(\rho, \mathbf{M}, \mathbf{N} )$ a strategy for $G$.
\begin{enumerate}
\item The register $AB$ is prepared in state $\rho$.
For $i = 1, 2, \ldots, n$, Alice applies the measurement
$\{ M_i^s \}$ to $A$ and records the result in a classical register $S_i$.

\item Referee gives Alice and Bob randomly chosen inputs $x \in \mathcal{X}$
and $y \in \mathcal{Y}$, respectively.

\item Alice returns the register $S_x$.  Bob measures $B$ with $\{ N_y^t \}_t$
and reports the result.
\end{enumerate}
}
}
\caption{A process in which Alice is forced to behave classically.}
\label{fig:fclass}
\end{figure}

\textit{Proof.} Let $Y = ( \Gamma, \mathbf{ M}, \mathbf{N}, \mathbf{P} )$ be a quantum strategy for 
$G_K$ on a space $A \otimes B \otimes E$. Let $\rho = \Gamma^{AB}$, and
for any $x \in \mathcal{X}, s \in \mathcal{S}$, let $\rho_{x}^{s}$ denote
the subnormalized state of $AB$ induced by the measurement $P_{xy}^{s}$ on
$E$.

For any $x, y$, the probability that Alice's and Eve's outputs will
disagree when the input is $(x, y, (x,y))$ is given by the quantity $(1 - \sum_s \Tr ( M_x^s \rho_{x}^s ))$,
which we denote by $\delta_x$.
Note that if the average failure probability $\sum_x \delta_x / n$ exceeds $1/K$, then (since a score of $-K$ is awarded
when Eve fails to guess Alice's output) the score achieved by $Y$ obviously
cannot exceed $0$.  So, we will assume for the remainder of the proof that
$\sum_x \delta_x/n \leq 1/K$.

By Proposition~\ref{undisturbprop} in the appendix, we have
\begin{eqnarray}
\left\| \sum_s M_x^s \rho M_x^s  - \rho \right\|_1 & \leq & 4 \sqrt{\delta_{x}}
\end{eqnarray}
for any $x,y$.  Therefore if we let $W_x$ denote the completely positive trace-preserving map
on $A$ given by $X \mapsto \sum_s M_x^s X M_x^s$, we obtain
the following distance inequalities for the states obtained
by applying the maps $W_x$ sequentially:
\begin{eqnarray}
&& \left\| W_i W_{i-1} \cdots W_1 ( \rho ) - \rho \right\|_1 \\
& \leq &
\nonumber \sum_{j=1}^{i} \left\| W_{i} W_{i-1} \cdots W_j ( \rho ) - W_{i-1} W_{i-2} \cdots W_{j+1} ( \rho ) \right\|_1 \\
\label{nonincstep}
& \leq & \sum_{j=1}^{i} \left\| W_j ( \rho ) - \rho \right\|_1 \leq \label{missbound} \sum_{j=1}^i 4 \sqrt{ \delta_j}.
\end{eqnarray}  
(Here we have used the fact that $\left\| \cdot \right\|_1$ is non-increasing under quantum processes.)

Observe that in the process in Figure~\ref{fig:fclass}, the state that Alice and Bob
measure at step 3 is separable, and so their expected score cannot exceed
$0$.  On the other hand, by (\ref{missbound}), the state of the register $AB$ is never more
than trace distance $\sum_{j=1}^n 4 \sqrt{ \delta_j}$ from the original state $\rho$,
and so the expected score achieved in Figure~\ref{fig:fclass} also cannot be
less than $\omega ( G, Y ) - \sum_{j=1}^n 4 \sqrt{ \delta_j}$.  Thus we have
\begin{eqnarray}
\omega ( G, Y ) & \leq & \sum_{j=1}^n 4 \sqrt{ \delta_j} 
\end{eqnarray}
which implies $\omega ( G, Y ) \leq 4 \sqrt{n} \sqrt{ \sum_{j=1}^n \delta_j }$. 
Since we have assumed $\sum_x \delta_x \leq n/K$, this yields
the desired result.  $\Box$

\paragraph*{\textbf{Security proof.}}

We will now prove the security of the protocol in Figure~\ref{fig:prot}
by considering the ``mirrored'' version of the protocol as
shown in Figure~\ref{fig:mirror}.

\begin{figure}
\fbox{\parbox{3.2in}{
\textit{Parameters:}
\begin{enumerate}
\item[-] A $2$-player Bell game $G$ with alphabet size $n$.
\item[-] A real constant $\delta > 0$ and positive integers $N, J$.
\item[-] A bipartite state $\Sigma$ of registers $AB$.
\end{enumerate}

\begin{enumerate}
\item Registers $ABA'B'$ are prepared in the canonical purification of the state $\Sigma$.

\item The referee prepares $n$-valued registers $X_1, \ldots, X_N, X'_1, \ldots, X'_N
Y_1, \ldots, Y_N, Y'_1, \ldots, Y'_N$, and
$D$-valued registers $F, F'$ (where $D$ denotes the hash family
from step~\ref{step:randext} in Figure~\ref{fig:prot})
so that for each register $Z$ the corresponding
primed register $Z'$ is in a Bell state with $Z$.  The
referee gives all primed registers to the adversary.

\item \label{genproc1} 
The referee measures the registers $\mathbf{X}, \mathbf{Y}$
in the standard bases to obtain $x_1, \ldots, x_N$ and $y_1, \ldots, y_N$,
which are given sequentially to Alice and Bob who return outputs
$s_1, \ldots, s_N, t_1, \ldots, t_N$. 

\item \label{genproc2} The referee checks whether the average score exceeds
$\delta$.  If not, the referee considers the protocol aborted.
If so, the referee measures $F$, and then computes $\textbf{V} := F ( \textbf{S} )$.

\item The adversary carries out step \ref{genproc1} above herself,
using the registers
$A', B', \mathbf{X}', \mathbf{Y}'$ and the conjugates of the measurements
used by Alice and Bob, to obtain outputs $\mathbf{S}', \mathbf{T}'$.  If the average score at $G$
is less than $\delta$,
the adversary considers the protocol aborted.  If not, she computes $\textbf{V}' := F' ( \textbf{S}' )$.
\end{enumerate}
}
}
\caption{The mirrored random number generation protocol.}
\label{fig:mirror}
\end{figure}

\begin{proposition}
For the process in Figure~\ref{fig:mirror}, let $succ$ and $succ'$ denote the events
that the referee and the adversary consider the protocol to have succeeded (respectively).  Then,
\begin{eqnarray}
\label{3events}
\mathbf{P} ( (\mathbf{S} = \mathbf{S}') \wedge succ \wedge succ' ) & \leq & 
e^{- \Omega ( N \delta^6 /n^4 )}.
\end{eqnarray}
\end{proposition}

\textit{Proof.}
For any $K \geq 1$, if the three events on the left side of (\ref{3events}) all occur, then Alice
and Bob and the adversary have achieved an average score of at least $\delta$ at
the repeated game $(G_K)^N$ using a sequential tripartite strategy.  By Propositions~\ref{prop:rep} and
\ref{prop:limiting}, the probability of such a score
is no more than
\begin{eqnarray}
\exp ( - N (\delta - 4 n / \sqrt{ K } )^2 / 8 K^2 ).
\end{eqnarray}
Setting $K = (8 n / \delta )^2$ yields the desired result. $\Box$

Note that the event $((\mathbf{V} = \mathbf{V'} ) \wedge succ \wedge succ')$
can occur only if either  $((\mathbf{S} = \mathbf{S'} ) \wedge succ \wedge succ')$
occurs, or if $((\mathbf{S} \neq \mathbf{S'} ) \wedge succ \wedge succ')$ occurs
but nonetheless $F ( \mathbf{S} ) = F ( \mathbf{S}')$.  Since $F$ is chosen
from a $2$-universal hash family, we have
\begin{eqnarray*}
&& \mathbf{P} ( (\mathbf{V} = \mathbf{V}' \wedge succ \wedge succ' ) \\
& \leq & 
e^{- \Omega ( N \delta^6 /n^4 )}  + 2^{-J} \mathbf{P} ( succ \wedge succ' ).
\end{eqnarray*}
By Proposition~\ref{prop:whatmirroris}, the register $\mathbf{V}'$ in Figure~\ref{fig:mirror}
is precisely the result of the adversary using a pretty good measurement in Figure~\ref{fig:prot} 
in order to guess $\mathbf{V}$.
Thus by Proposition~\ref{prop:pgp} (with $C = \mathbf{V}$ and $Q = \mathbf{XY}FE$), we obtain the following.
\begin{theorem}
\label{mainthm}
Let $\rho$ denote the final state of the registers in Figure~\ref{fig:prot}.  Then,
\begin{eqnarray*}
\label{finalexpr}
\left\| \rho_{succ}^{\mathbf{VXY} F E} - U_\mathbf{V} \otimes \rho_{succ}^{\mathbf{XY} F E} \right\|_1
& \leq & 2^{J/2 - \Omega ( N \delta^6 / n^4 )}. \hskip0.2in \Box
\label{finalexpr}
\end{eqnarray*}
\end{theorem}

Note that if we fix $\delta, n$ and let $J = \lfloor c N \rfloor$ for some
sufficiently small $c > 0$, the expression on the right of the inequality above
vanishes exponentially.  Thus
random number generation with a linear rate and negligible error term is achieved.

\appendix

\section{Supplementary Proofs}

\subsection{The proof of Proposition~\ref{prop:pgp}}

\label{app:pgp}

We follow the proof of Lemma 4 in \cite{Tomamichel:2011}.  
Let $X = \alpha^Q$ and $Y = \alpha^{QC}_{succ}$.  Note that $\Tr ( X ) = 1$,
and therefore $\left\| X^{1/d} \right\|_d = 1$ for any $d$.  
By Holder's inequality, we have the following.
\begin{eqnarray*}
&& \left\| Y - Y^Q \otimes U_C \right\|_1 \\
& \leq
& \left\| X^{1/4} \otimes I_C \right\|_4 \left\| X^{-1/4} (Y - Y^Q \otimes U_C) X^{-1/4} \right\|_2 \\ && \cdot \left\| X^{1/4} \otimes I_C \right\|_4 \\
& = & \left| C \right|^{1/4} \cdot \Tr \left[ \left( X^{-1/4} (Y - Y^Q \otimes U_C) X^{-1/4} \right)^2 \right]^{1/2} \left| C \right|^{1/4} \\
& = & \left| C \right|^{1/2} \left\{ \Tr \left[ \left( X^{-1/4} Y  X^{-1/4} \right)^2 \right] \right. \\
&& - 2 \Tr \left[ X^{-1/2} Y X^{-1/2} ( Y^Q \otimes U_C ) X^{-1/4} \right] \\
&& \left. +  \Tr \left[ \left( X^{-1/4} (Y^Q \otimes U_C )  X^{-1/4} \right)^2 \right] \right\}^{1/2} \\
& = & \left| C \right|^{1/2} \left\{ \Tr \left[ \left( X^{-1/4} Y  X^{-1/4} \right)^2 \right] \right. \\
&& \left. -  \frac{1}{\left| C \right|} \Tr \left[ \left( X^{-1/4} (Y^Q )  X^{-1/4} \right)^2 \right] \right\}^{1/2},
\end{eqnarray*}
where we have used the fact that $\Tr [ (Y^Q \otimes U_C ) Z] = \frac{1}{\left| C \right| } \Tr [ Y^Q Z^Q]$
for any Hermitian operator $Z$ on $QC$. By substitution we obtain the desired result.

\subsection{Predictable measurements}  We reprove a result used by other authors \cite{Kempe:2011,
Winter:thesis}.  The following proposition asserts that if a quantum-classical
state of a register $QC$ is such that $C$ can be accurately guessed
from a measurement on $Q$, then that same measurement does
not disturb the state by much.  

\begin{proposition}
\label{undisturbprop}
Let $QC$ be a classical quantum register in state $\alpha$,
and let $\{ P^c \}_c$ be a projective measurement on $Q$ whose outcome
agrees with $C$ with probability $1 - \delta$.  Then,
\begin{eqnarray}
\label{keyineq}
\left\| \sum_{c \in C} P^c \alpha P^c - \alpha \right\|_1 & \leq & 4 \sqrt{\delta}.
\end{eqnarray}
\end{proposition}

\textit{Proof.} Our proof is similar to that of \cite{Winter:thesis}, Lemma I.4.
First suppose that $\alpha$ is concentrated on a single basic state
of $C$, i.e., $P_\alpha ( C = z ) = 1$ for some $z$.  Then,
\begin{eqnarray*}
\Tr ( (P^z) \alpha ) = 1 - \delta,
\end{eqnarray*}
and therefore
\begin{eqnarray*}
&& \left\| P^z \alpha P^z  - \alpha \right\|_1 \\
& = &
\left\| (P^z)^\perp \alpha P^z + (P^z ) \alpha (P^z )^\perp + (P^z ) \alpha (P^z )^\perp \right\|_1  \\
& \leq  & \left\| (P^z)^\perp \alpha P^z \left\|_1 + \right\| (P^z ) \alpha (P^z )^\perp \right\|_1 + \left\| (P^z )^\perp \alpha (P^z )^\perp \right\|_1  \\
& = & 2 \left\| (P^z)^\perp \alpha P^z \right\|_1 + \delta  \\
& \leq  & 2 \left\| (P^z)^\perp \sqrt{\alpha } \right\|_2 \left\| \sqrt{\alpha}  P^z \right\|_2 + \delta  \\
& \leq  & 2 \sqrt{ \left\| (P^z)^\perp \alpha (P^z)^{\perp} \right\|_1 } \sqrt{ \left\| P^z \alpha  P^z \right\|_1} +\delta   \\
& =  & 2 \sqrt{ ( 1- \delta) \delta } + \delta  \\
& \leq  & 3 \sqrt{\delta}.
\end{eqnarray*}
And, $\left\| P^z \alpha P^z - \sum_{c} P^c \alpha P^c \right\|_1 \leq \delta \leq \sqrt{\delta}$ which yields the desired
result.

The general case now follows, since any state of $CQ$ is a convex combination of states that are concentrated
on a single value of $C$, the function $\left\| \cdot \right\|_1$ is convex,
and the square root function is concave. $\Box$

\subsection{Two-universal hash families}

\label{subsec:hash}

We make use of some standard ideas (see, e.g., 
section 4.6.1 in \cite{Katz2015}).
Let $P, R$ be finite sets with $\left| R \right| \leq \left| P \right|$.
Then, a set of
functions $D$ from $P$ to $R$ is
\textit{$2$-universal} if for any distinct $p, p' \in P$,
the probability that a function $F$ chosen uniformly at random
from $D$ will satisfy $F ( p ) = F ( p' )$ is less than or
equal to $1 / \left| R \right|$. 

\begin{proposition}
\label{prop:hash}
Let $P$ be a finite set and let $u$ be a positive
integer with $2^u \leq \left| P \right|$. Then
there exists a $2$-universal set of functions
from $P$ to $\mathbb{F}_2^u$ of size $\leq 4 \left| P \right|^2$. 
\end{proposition}

\textit{Proof.}  Let $v$ be such that $2^{v-1} < \left| P \right| \leq 2^v$.  
Without loss of generality, we may assume
that $P \subseteq \mathbb{F}_{2^v}$.  Let $D'$
be the set of all affine endomorphisms ($X \mapsto
aX + b$) of $\mathbb{F}_{2^v}$.  Fix a function
$T \colon \mathbb{F}_{2^v} \to \mathbb{F}_2^u$ such that each element
of $R$ has exactly $2^{u-v}$ pre-images, and
let $D =  T \circ D'$.  Note that $\left| D \right| \leq \left| D' \right| = (2^v)^2
\leq 4 \left| P \right|^2$.

For any $p \neq p'$ and $q, q'$ in $\mathbb{F}_{2^v}$, there
is exactly one function in $D'$ which maps $(p, p')$
to $(q, q')$.  Thus the distribution of $(F ( p ) , F ( p' ) )$
on $P \times P$ is uniform when $F$ is chosen at
random from $D'$, and likewise $(T \circ F ( p ) , T \circ F ( p' ))$
is uniform on $R \times R$.  The desired result follows.

\vskip0.6in

\bibliography{Bell}

\begin{thebibliography}{23}%
\makeatletter
\providecommand \@ifxundefined [1]{%
 \@ifx{#1\undefined}
}%
\providecommand \@ifnum [1]{%
 \ifnum #1\expandafter \@firstoftwo
 \else \expandafter \@secondoftwo
 \fi
}%
\providecommand \@ifx [1]{%
 \ifx #1\expandafter \@firstoftwo
 \else \expandafter \@secondoftwo
 \fi
}%
\providecommand \natexlab [1]{#1}%
\providecommand \enquote  [1]{``#1''}%
\providecommand \bibnamefont  [1]{#1}%
\providecommand \bibfnamefont [1]{#1}%
\providecommand \citenamefont [1]{#1}%
\providecommand \href@noop [0]{\@secondoftwo}%
\providecommand \href [0]{\begingroup \@sanitize@url \@href}%
\providecommand \@href[1]{\@@startlink{#1}\@@href}%
\providecommand \@@href[1]{\endgroup#1\@@endlink}%
\providecommand \@sanitize@url [0]{\catcode `\\12\catcode `\$12\catcode
  `\&12\catcode `\#12\catcode `\^12\catcode `\_12\catcode `\%12\relax}%
\providecommand \@@startlink[1]{}%
\providecommand \@@endlink[0]{}%
\providecommand \url  [0]{\begingroup\@sanitize@url \@url }%
\providecommand \@url [1]{\endgroup\@href {#1}{\urlprefix }}%
\providecommand \urlprefix  [0]{URL }%
\providecommand \Eprint [0]{\href }%
\providecommand \doibase [0]{http://dx.doi.org/}%
\providecommand \selectlanguage [0]{\@gobble}%
\providecommand \bibinfo  [0]{\@secondoftwo}%
\providecommand \bibfield  [0]{\@secondoftwo}%
\providecommand \translation [1]{[#1]}%
\providecommand \BibitemOpen [0]{}%
\providecommand \bibitemStop [0]{}%
\providecommand \bibitemNoStop [0]{.\EOS\space}%
\providecommand \EOS [0]{\spacefactor3000\relax}%
\providecommand \BibitemShut  [1]{\csname bibitem#1\endcsname}%
\let\auto@bib@innerbib\@empty
\bibitem [{\citenamefont {Colbeck}(2007)}]{colbeck2007quantum}%
  \BibitemOpen
  \bibfield  {author} {\bibinfo {author} {\bibfnamefont {R.}~\bibnamefont
  {Colbeck}},\ }\href@noop {} {\enquote {\bibinfo {title} {Quantum and
  relativistic protocols for secure multi-party computation},}\ }\bibinfo
  {howpublished} {{Ph.D.} thesis, University of York, arXiv:0911.3814}
  (\bibinfo {year} {2007})\BibitemShut {NoStop}%
\bibitem [{\citenamefont {Colbeck}\ and\ \citenamefont
  {Renner}(2012)}]{Colbeck:2012}%
  \BibitemOpen
  \bibfield  {author} {\bibinfo {author} {\bibfnamefont {R.}~\bibnamefont
  {Colbeck}}\ and\ \bibinfo {author} {\bibfnamefont {R.}~\bibnamefont
  {Renner}},\ }\href@noop {} {\bibfield  {journal} {\bibinfo  {journal} {Nature
  Physics}\ }\textbf {\bibinfo {volume} {8}},\ \bibinfo {pages} {450} (\bibinfo
  {year} {2012})}\BibitemShut {NoStop}%
\bibitem [{\citenamefont {Ekert}(1991)}]{Ekert:1991}%
  \BibitemOpen
  \bibfield  {author} {\bibinfo {author} {\bibfnamefont {A.~K.}\ \bibnamefont
  {Ekert}},\ }\href {\doibase 10.1103/PhysRevLett.67.661} {\bibfield  {journal}
  {\bibinfo  {journal} {Phys. Rev. Lett.}\ }\textbf {\bibinfo {volume} {67}},\
  \bibinfo {pages} {661} (\bibinfo {year} {1991})}\BibitemShut {NoStop}%
\bibitem [{\citenamefont {Silman}\ \emph {et~al.}(2011)\citenamefont {Silman},
  \citenamefont {Chailloux}, \citenamefont {Aharon}, \citenamefont {Kerenidis},
  \citenamefont {Pironio},\ and\ \citenamefont {Massar}}]{Silman:2011}%
  \BibitemOpen
  \bibfield  {author} {\bibinfo {author} {\bibfnamefont {J.}~\bibnamefont
  {Silman}}, \bibinfo {author} {\bibfnamefont {A.}~\bibnamefont {Chailloux}},
  \bibinfo {author} {\bibfnamefont {N.}~\bibnamefont {Aharon}}, \bibinfo
  {author} {\bibfnamefont {I.}~\bibnamefont {Kerenidis}}, \bibinfo {author}
  {\bibfnamefont {S.}~\bibnamefont {Pironio}}, \ and\ \bibinfo {author}
  {\bibfnamefont {S.}~\bibnamefont {Massar}},\ }\href {\doibase
  10.1103/PhysRevLett.106.220501} {\bibfield  {journal} {\bibinfo  {journal}
  {Phys. Rev. Lett.}\ }\textbf {\bibinfo {volume} {106}},\ \bibinfo {pages}
  {220501} (\bibinfo {year} {2011})}\BibitemShut {NoStop}%
\bibitem [{\citenamefont {Pironio}\ \emph {et~al.}(2010)\citenamefont
  {Pironio}, \citenamefont {Ac{\'\i}n}, \citenamefont {Massar}, \citenamefont
  {de~La~Giroday}, \citenamefont {Matsukevich}, \citenamefont {Maunz},
  \citenamefont {Olmschenk}, \citenamefont {Hayes}, \citenamefont {Luo},
  \citenamefont {Manning} \emph {et~al.}}]{pironio2010random}%
  \BibitemOpen
  \bibfield  {author} {\bibinfo {author} {\bibfnamefont {S.}~\bibnamefont
  {Pironio}}, \bibinfo {author} {\bibfnamefont {A.}~\bibnamefont {Ac{\'\i}n}},
  \bibinfo {author} {\bibfnamefont {S.}~\bibnamefont {Massar}}, \bibinfo
  {author} {\bibfnamefont {A.~B.}\ \bibnamefont {de~La~Giroday}}, \bibinfo
  {author} {\bibfnamefont {D.~N.}\ \bibnamefont {Matsukevich}}, \bibinfo
  {author} {\bibfnamefont {P.}~\bibnamefont {Maunz}}, \bibinfo {author}
  {\bibfnamefont {S.}~\bibnamefont {Olmschenk}}, \bibinfo {author}
  {\bibfnamefont {D.}~\bibnamefont {Hayes}}, \bibinfo {author} {\bibfnamefont
  {L.}~\bibnamefont {Luo}}, \bibinfo {author} {\bibfnamefont {T.~A.}\
  \bibnamefont {Manning}},  \emph {et~al.},\ }\href@noop {} {\bibfield
  {journal} {\bibinfo  {journal} {Nature}\ }\textbf {\bibinfo {volume} {464}},\
  \bibinfo {pages} {1021} (\bibinfo {year} {2010})}\BibitemShut {NoStop}%
\bibitem [{\citenamefont {Bierhorst}\ \emph {et~al.}(2018)\citenamefont
  {Bierhorst}, \citenamefont {Knill}, \citenamefont {Glancy}, \citenamefont
  {Zhang}, \citenamefont {Mink}, \citenamefont {Jordan}, \citenamefont
  {Rommal}, \citenamefont {Liu}, \citenamefont {Christensen}, \citenamefont
  {Nam} \emph {et~al.}}]{bierhorst2018experimentally}%
  \BibitemOpen
  \bibfield  {author} {\bibinfo {author} {\bibfnamefont {P.}~\bibnamefont
  {Bierhorst}}, \bibinfo {author} {\bibfnamefont {E.}~\bibnamefont {Knill}},
  \bibinfo {author} {\bibfnamefont {S.}~\bibnamefont {Glancy}}, \bibinfo
  {author} {\bibfnamefont {Y.}~\bibnamefont {Zhang}}, \bibinfo {author}
  {\bibfnamefont {A.}~\bibnamefont {Mink}}, \bibinfo {author} {\bibfnamefont
  {S.}~\bibnamefont {Jordan}}, \bibinfo {author} {\bibfnamefont
  {A.}~\bibnamefont {Rommal}}, \bibinfo {author} {\bibfnamefont {Y.-K.}\
  \bibnamefont {Liu}}, \bibinfo {author} {\bibfnamefont {B.}~\bibnamefont
  {Christensen}}, \bibinfo {author} {\bibfnamefont {S.~W.}\ \bibnamefont
  {Nam}},  \emph {et~al.},\ }\href@noop {} {\bibfield  {journal} {\bibinfo
  {journal} {Nature}\ }\textbf {\bibinfo {volume} {556}},\ \bibinfo {pages}
  {223} (\bibinfo {year} {2018})}\BibitemShut {NoStop}%
\bibitem [{\citenamefont {Fehr}\ \emph {et~al.}(2013)\citenamefont {Fehr},
  \citenamefont {Gelles},\ and\ \citenamefont {Schaffner}}]{fehr2013security}%
  \BibitemOpen
  \bibfield  {author} {\bibinfo {author} {\bibfnamefont {S.}~\bibnamefont
  {Fehr}}, \bibinfo {author} {\bibfnamefont {R.}~\bibnamefont {Gelles}}, \ and\
  \bibinfo {author} {\bibfnamefont {C.}~\bibnamefont {Schaffner}},\ }\href@noop
  {} {\bibfield  {journal} {\bibinfo  {journal} {Physical Review A}\ }\textbf
  {\bibinfo {volume} {87}},\ \bibinfo {pages} {012335} (\bibinfo {year}
  {2013})}\BibitemShut {NoStop}%
\bibitem [{\citenamefont {Pironio}\ and\ \citenamefont
  {Massar}(2013)}]{pironio2013security}%
  \BibitemOpen
  \bibfield  {author} {\bibinfo {author} {\bibfnamefont {S.}~\bibnamefont
  {Pironio}}\ and\ \bibinfo {author} {\bibfnamefont {S.}~\bibnamefont
  {Massar}},\ }\href@noop {} {\bibfield  {journal} {\bibinfo  {journal}
  {Physical Review A}\ }\textbf {\bibinfo {volume} {87}},\ \bibinfo {pages}
  {012336} (\bibinfo {year} {2013})}\BibitemShut {NoStop}%
\bibitem [{\citenamefont {DiVincenzo}\ \emph {et~al.}(2004)\citenamefont
  {DiVincenzo}, \citenamefont {Horodecki}, \citenamefont {Leung}, \citenamefont
  {Smolin},\ and\ \citenamefont {Terhal}}]{DiV:2004}%
  \BibitemOpen
  \bibfield  {author} {\bibinfo {author} {\bibfnamefont {D.~P.}\ \bibnamefont
  {DiVincenzo}}, \bibinfo {author} {\bibfnamefont {M.}~\bibnamefont
  {Horodecki}}, \bibinfo {author} {\bibfnamefont {D.~W.}\ \bibnamefont
  {Leung}}, \bibinfo {author} {\bibfnamefont {J.~A.}\ \bibnamefont {Smolin}}, \
  and\ \bibinfo {author} {\bibfnamefont {B.~M.}\ \bibnamefont {Terhal}},\
  }\href {\doibase 10.1103/PhysRevLett.92.067902} {\bibfield  {journal}
  {\bibinfo  {journal} {Phys. Rev. Lett.}\ }\textbf {\bibinfo {volume} {92}},\
  \bibinfo {pages} {067902} (\bibinfo {year} {2004})}\BibitemShut {NoStop}%
\bibitem [{\citenamefont {Vazirani}\ and\ \citenamefont
  {Vidick}(2012)}]{vazirani2012}%
  \BibitemOpen
  \bibfield  {author} {\bibinfo {author} {\bibfnamefont {U.}~\bibnamefont
  {Vazirani}}\ and\ \bibinfo {author} {\bibfnamefont {T.}~\bibnamefont
  {Vidick}},\ }in\ \href@noop {} {\emph {\bibinfo {booktitle} {Proceedings of
  the forty-fourth annual ACM Symposium on Theory of Computing (STOC)}}}\
  (\bibinfo {year} {2012})\ pp.\ \bibinfo {pages} {61--76}\BibitemShut
  {NoStop}%
\bibitem [{\citenamefont {Vazirani}\ and\ \citenamefont
  {Vidick}(2014)}]{vazirani2014fully}%
  \BibitemOpen
  \bibfield  {author} {\bibinfo {author} {\bibfnamefont {U.}~\bibnamefont
  {Vazirani}}\ and\ \bibinfo {author} {\bibfnamefont {T.}~\bibnamefont
  {Vidick}},\ }\href@noop {} {\bibfield  {journal} {\bibinfo  {journal}
  {Physical review letters}\ }\textbf {\bibinfo {volume} {113}},\ \bibinfo
  {pages} {140501} (\bibinfo {year} {2014})}\BibitemShut {NoStop}%
\bibitem [{\citenamefont {Miller}\ and\ \citenamefont {Shi}(2016)}]{MY14-1}%
  \BibitemOpen
  \bibfield  {author} {\bibinfo {author} {\bibfnamefont {C.~A.}\ \bibnamefont
  {Miller}}\ and\ \bibinfo {author} {\bibfnamefont {Y.}~\bibnamefont {Shi}},\
  }\href {\doibase 10.1145/2885493} {\bibfield  {journal} {\bibinfo  {journal}
  {J. ACM}\ }\textbf {\bibinfo {volume} {63}},\ \bibinfo {pages} {33:1}
  (\bibinfo {year} {2016})}\BibitemShut {NoStop}%
\bibitem [{\citenamefont {Miller}\ and\ \citenamefont
  {Shi}(2017)}]{miller2017universal}%
  \BibitemOpen
  \bibfield  {author} {\bibinfo {author} {\bibfnamefont {C.~A.}\ \bibnamefont
  {Miller}}\ and\ \bibinfo {author} {\bibfnamefont {Y.}~\bibnamefont {Shi}},\
  }\href@noop {} {\bibfield  {journal} {\bibinfo  {journal} {SIAM Journal on
  Computing}\ }\textbf {\bibinfo {volume} {46}},\ \bibinfo {pages} {1304}
  (\bibinfo {year} {2017})}\BibitemShut {NoStop}%
\bibitem [{\citenamefont {Dupuis}\ \emph {et~al.}(2016)\citenamefont {Dupuis},
  \citenamefont {Fawzi},\ and\ \citenamefont {Renner}}]{Dupuis:2016}%
  \BibitemOpen
  \bibfield  {author} {\bibinfo {author} {\bibfnamefont {F.}~\bibnamefont
  {Dupuis}}, \bibinfo {author} {\bibfnamefont {O.}~\bibnamefont {Fawzi}}, \
  and\ \bibinfo {author} {\bibfnamefont {R.}~\bibnamefont {Renner}},\
  }\href@noop {} {\enquote {\bibinfo {title} {Entropy accumulation},}\ }
  (\bibinfo {year} {2016}),\ \bibinfo {note} {arXiv:1607.01796}\BibitemShut
  {NoStop}%
\bibitem [{\citenamefont {Arnon-Friedman}\ \emph {et~al.}(2016)\citenamefont
  {Arnon-Friedman}, \citenamefont {Renner},\ and\ \citenamefont
  {Vidick}}]{Arnon:2016}%
  \BibitemOpen
  \bibfield  {author} {\bibinfo {author} {\bibfnamefont {R.}~\bibnamefont
  {Arnon-Friedman}}, \bibinfo {author} {\bibfnamefont {R.}~\bibnamefont
  {Renner}}, \ and\ \bibinfo {author} {\bibfnamefont {T.}~\bibnamefont
  {Vidick}},\ }\href@noop {} {\enquote {\bibinfo {title} {Simple and tight
  device-independent security proofs},}\ } (\bibinfo {year} {2016}),\ \bibinfo
  {note} {arXiv:1607.01797}\BibitemShut {NoStop}%
\bibitem [{\citenamefont {Arnon-Friedman}\ \emph {et~al.}(2018)\citenamefont
  {Arnon-Friedman}, \citenamefont {Dupuis}, \citenamefont {Fawzi},
  \citenamefont {Renner},\ and\ \citenamefont {Vidick}}]{arnon2018practical}%
  \BibitemOpen
  \bibfield  {author} {\bibinfo {author} {\bibfnamefont {R.}~\bibnamefont
  {Arnon-Friedman}}, \bibinfo {author} {\bibfnamefont {F.}~\bibnamefont
  {Dupuis}}, \bibinfo {author} {\bibfnamefont {O.}~\bibnamefont {Fawzi}},
  \bibinfo {author} {\bibfnamefont {R.}~\bibnamefont {Renner}}, \ and\ \bibinfo
  {author} {\bibfnamefont {T.}~\bibnamefont {Vidick}},\ }\href@noop {}
  {\bibfield  {journal} {\bibinfo  {journal} {Nature communications}\ }\textbf
  {\bibinfo {volume} {9}},\ \bibinfo {pages} {459} (\bibinfo {year}
  {2018})}\BibitemShut {NoStop}%
\bibitem [{\citenamefont {Jain}\ \emph {et~al.}(2017)\citenamefont {Jain},
  \citenamefont {Miller},\ and\ \citenamefont {Shi}}]{Jain:2017}%
  \BibitemOpen
  \bibfield  {author} {\bibinfo {author} {\bibfnamefont {R.}~\bibnamefont
  {Jain}}, \bibinfo {author} {\bibfnamefont {C.~A.}\ \bibnamefont {Miller}}, \
  and\ \bibinfo {author} {\bibfnamefont {Y.}~\bibnamefont {Shi}},\ }\href@noop
  {} {\enquote {\bibinfo {title} {Parallel device-independent quantum key
  distribution},}\ } (\bibinfo {year} {2017}),\ \bibinfo {note}
  {{arXiv:1703.05426}}\BibitemShut {NoStop}%
\bibitem [{\citenamefont {Tomamichel}\ \emph {et~al.}(2011)\citenamefont
  {Tomamichel}, \citenamefont {Schaffner}, \citenamefont {Smith},\ and\
  \citenamefont {Renner}}]{Tomamichel:2011}%
  \BibitemOpen
  \bibfield  {author} {\bibinfo {author} {\bibfnamefont {M.}~\bibnamefont
  {Tomamichel}}, \bibinfo {author} {\bibfnamefont {C.}~\bibnamefont
  {Schaffner}}, \bibinfo {author} {\bibfnamefont {A.}~\bibnamefont {Smith}}, \
  and\ \bibinfo {author} {\bibfnamefont {R.}~\bibnamefont {Renner}},\
  }\href@noop {} {\bibfield  {journal} {\bibinfo  {journal} {IEEE Transactions
  on Information Theory}\ }\textbf {\bibinfo {volume} {57}},\ \bibinfo {pages}
  {5524} (\bibinfo {year} {2011})}\BibitemShut {NoStop}%
\bibitem [{\citenamefont {Kempe}\ \emph {et~al.}(2011)\citenamefont {Kempe},
  \citenamefont {Kobayashi}, \citenamefont {Matsumoto}, \citenamefont {Toner},\
  and\ \citenamefont {Vidick}}]{Kempe:2011}%
  \BibitemOpen
  \bibfield  {author} {\bibinfo {author} {\bibfnamefont {J.}~\bibnamefont
  {Kempe}}, \bibinfo {author} {\bibfnamefont {H.}~\bibnamefont {Kobayashi}},
  \bibinfo {author} {\bibfnamefont {K.}~\bibnamefont {Matsumoto}}, \bibinfo
  {author} {\bibfnamefont {B.}~\bibnamefont {Toner}}, \ and\ \bibinfo {author}
  {\bibfnamefont {T.}~\bibnamefont {Vidick}},\ }\href@noop {} {\bibfield
  {journal} {\bibinfo  {journal} {SIAM Journal on Computing}\ }\textbf
  {\bibinfo {volume} {40}},\ \bibinfo {pages} {848} (\bibinfo {year}
  {2011})}\BibitemShut {NoStop}%
\bibitem [{\citenamefont {Alon}\ and\ \citenamefont
  {Spencer}(2004)}]{Alon:2004}%
  \BibitemOpen
  \bibfield  {author} {\bibinfo {author} {\bibfnamefont {N.}~\bibnamefont
  {Alon}}\ and\ \bibinfo {author} {\bibfnamefont {J.~H.}\ \bibnamefont
  {Spencer}},\ }\href@noop {} {\emph {\bibinfo {title} {The probabilistic
  method}}}\ (\bibinfo  {publisher} {John Wiley \& Sons},\ \bibinfo {year}
  {2004})\BibitemShut {NoStop}%
\bibitem [{\citenamefont {Bhatia}(1997)}]{Bhatia:1997}%
  \BibitemOpen
  \bibfield  {author} {\bibinfo {author} {\bibfnamefont {R.}~\bibnamefont
  {Bhatia}},\ }\href@noop {} {\emph {\bibinfo {title} {Matrix Analysis}}}\
  (\bibinfo  {publisher} {Springer-Verlag},\ \bibinfo {year}
  {1997})\BibitemShut {NoStop}%
\bibitem [{\citenamefont {Winter}(1999)}]{Winter:thesis}%
  \BibitemOpen
  \bibfield  {author} {\bibinfo {author} {\bibfnamefont {A.}~\bibnamefont
  {Winter}},\ }\emph {\bibinfo {title} {Coding Theorems of Quantum Information
  Theory}},\ \href@noop {} {Ph.D. thesis},\ \bibinfo  {school} {Universitat
  Bielfeld} (\bibinfo {year} {1999})\BibitemShut {NoStop}%
\bibitem [{\citenamefont {Katz}\ and\ \citenamefont
  {Lindell}(2015)}]{Katz2015}%
  \BibitemOpen
  \bibfield  {author} {\bibinfo {author} {\bibfnamefont {J.}~\bibnamefont
  {Katz}}\ and\ \bibinfo {author} {\bibfnamefont {Y.}~\bibnamefont {Lindell}},\
  }\href@noop {} {\emph {\bibinfo {title} {Introduction to Modern
  Cryptography}}}\ (\bibinfo  {publisher} {CRC Press},\ \bibinfo {year}
  {2015})\BibitemShut {NoStop}%
\end{thebibliography}%

\end{document}